\documentclass{article}

\usepackage{amsmath, amsfonts, amssymb, amsthm}

\usepackage{mathtools}

\usepackage{caption}
\usepackage{subcaption}

\usepackage{scalerel}
\usepackage{subfig}

\usepackage{color}
\newcommand{\lca}{\ensuremath{\operatorname{lca}}}
\newcommand{\eT}{\ensuremath{\widehat{R}_o}}
\newcommand{\eP}{\ensuremath{\widehat{R}_p}}
\newcommand{\eX}{\ensuremath{\widehat{R}_x}}

\newcommand{\Gen}{\ensuremath{\mathbb{G}}}
\newcommand{\Spe}{\ensuremath{\mathbb{S}}}
\newcommand{\rt}[1]{\ensuremath{\mathsf{#1}}}

\newcommand{\ev}{\ensuremath{\stackrel{\hstretch{1.6}{\vstretch{0.3}{\wedge}}}{=}_t}}

\newtheorem{theorem}{Theorem}%[section]
\newtheorem{remark}{Remark}%[section]

\makeindex             % used for the subject index
\begin{document}

\title{Construction of Gene and Species Trees from Sequence Data incl.\ Orthologs, Paralogs, and Xenologs}

\author{Marc Hellmuth\\ University of Greifswald \\ Department of Mathematics and Computer Science \\ Walther-
           Rathenau-Strasse 47, \\ D-17487 Greifswald, Germany,  \\and \\
				Saarland University \\Center for Bioinformatics \\ Building E 2.1, 
				P.O.\ Box 151150, \\ D-66041 Saarbr{\"u}cken, Germany \\
			 Email: \texttt{mhellmuth@mailbox.org} 
				\and  Nicolas Wieseke 
	Leipzig University \\ Parallel Computing and Complex Systems Group \\ Department of 
     Computer Science \\ Augustusplatz 10, D-04109 Leipzig, Germany \\
    Email: \texttt	{wieseke@informatik.uni-leipzig.de}
	}

\date{\ }

\maketitle

\abstract{ 
Phylogenetic reconstruction aims at finding plausible hypotheses of the
evolutionary history of genes or species based on genomic sequence information.
The distinction of orthologous genes (genes that having a common ancestry and
diverged after a speciation) is crucial and lies at the heart of many genomic
studies. However, existing methods that rely only on 1:1 orthologs to infer
species trees are strongly restricted to a small set of allowed genes that
provide information about the species tree. The use of larger gene sets that
consist in addition of non-orthologous genes (e.g. so-called paralogous or
xenologous genes) considerably increases the information about the evolutionary
history of the respective species. 
In this work, we introduce a novel method to compute
species phylogenies based on sequence data including orthologs, paralogs or
even xenologs. 
}

\sloppy

\section{Introduction}
\label{sec:intro}

 Sequence-based phylogenetic approaches heavily rely
on initial data sets to be composed of 1:1 orthologous sequences only.
To this end alignments of protein or DNA
sequences are employed whose evolutionary history is believed
to be congruent to that of the respective species, a  property
that can be ensured most easily in the absence of gene duplications or
horizontal gene transfer.
Phylogenetic studies thus judiciously select families of genes that
rarely exhibit duplications (such as rRNAs, most ribosomal proteins,
and many of the housekeeping enzymes). 
In the presence of gene duplications, however, it becomes
necessary to distinguish between the evolutionary history of genes
(gene trees) and the evolutionary history of the species (species
trees) in which these genes reside.

Recent advances in mathematical phylogenetics, 
based on the theory of symbolic ultrametrics \cite{Boeckner:98},
have indicated that gene duplications can also convey
meaningful phylogenetic information provided orthologs and paralogs
can be distinguished with a degree of certainty \cite{HHH+13,HLS+15,HHH+12}.

Here, we examine a novel approach and explain the conceptional steps for the 
inference of species trees based on the knowledge of orthologs, paralogs or
even xenologs \cite{HHH+13, HLS+15,HHH+12}.

\section{Preliminaries}
\label{sec:prelim}

We give here a brief summary of the main definitions and concepts that are needed.

\paragraph{\textbf{Graphs, Gene Trees and Species Trees}}

An \emph{(undirected) graph} $G$ is a pair $(V,E)$ with non-empty vertex
set $V$ and edge set $E$ containing two-element subsets of $V$. 
A class of graphs that will play an important role in this contribution
are cographs. A graph $G=(V,E)$ is a \emph{cograph} iff $G$ does not contain
an induced path on four vertices, see \cite{Corneil:81,Corneil:85} for more details.

A \emph{tree} $T=(V,E)$ is a connected, cycle-free graph. We distinguish two types
of vertices in a tree: the \emph{leaves} which are contained in only one edge
and the \emph{inner} vertices which are contained in at least two edges. 
In order to avoid uninteresting trivial cases, we will usually assume
that $T$ has at least three leaves. 

A \emph{rooted} tree is a tree in which one special (inner) vertex is selected
to be the root. The \emph{least common ancestor} $\lca_T(x,y)$ of two vertices $x$ and $y$ 
in a rooted tree $T$ is the first (unique) vertex that lies on the path from $x$ to the
root and $y$ to the root. We say that a tree $T$ contains the triple $\rt{xy|z}$
if $x,y,$ and $z$ are leaves of $T$ and the path from $x$ to $y$ does
not intersect the path from $z$ to the root of $T$. A set of triples $\mathcal R$
is consistent if there is a rooted tree that contains all triples in $\mathcal R$.

An \emph{event-labeled} tree, usually denoted by the pair $(T,t)$, 
 is a rooted tree $T$ together with a 
map $t:V\to M$ that assigns to each inner vertex an event $m\in M$. 
For two leaves $x$ and $y$ of an event-labeled tree $(T,t)$ 
its least common ancestor
$\lca_T(x,y)$ is therefore marked with an event $t(\lca_T(x,y))=m$, 
which we denote for simplicity by $\lca_T(x,y)\ev m$.

In what follows, the set  $\Spe$ will always denote a set of species and
the set $\Gen$ a set of genes. We write $x\in X$ if a gene $x\in \Gen$ resides in 
the species $X\in \Spe$.

A \emph{species tree} (for $\Spe$) is a rooted tree $T$
with leaf-set $\Spe$.
A \emph{gene tree} (for $\Gen$)
is an event-labeled tree $(T,t)$ that has as leaf-set $\Gen$.

We refer the reader to \cite{sem-ste-03a} for an 
overview and important results on phylogenetics.

\paragraph{\textbf{Binary Relations and its Graph- and Tree-Representations}}

A \emph{(binary) relation} $R$ over (an underlying set) $\Gen$ is a subset 
of $\Gen \times \Gen$. We will write 
$\lfloor\Gen\times \Gen\rfloor_{\textrm{irr}}  \coloneqq  
 (\Gen\times \Gen)\setminus \{(x,x)\mid x\in \Gen\}$
to denote the irreflexive part of $\Gen\times \Gen$.

Each relation $R$ has a natural representation
as a graph $G_R=(\Gen, E_R)$ with vertex set $\Gen$ and edges 
connecting two vertices whenever they are in relation $R$.
In what follows, we will always deal with \emph{irreflexive symmetric} relations, 
which we call for simplicity just relations. Therefore, 
the corresponding graphs $G_R$ can be considered as undirected graphs without loops, 
that is, $\{x\}\not\in E_R$ and, additionally, $\{x,y\}\in E_R$ iff $(x,y)\in R$ (and thus, $(y,x)\in R$).

While Graph-Representations $G_R$ of $R$ are straightforward and defined 
for all binary relations, tree-representations of $R$ are a bit 
more difficile to derive and, even more annoying,
not every binary relation does have
a tree-representation. For each tree representing a relation $R$ over $\Gen$
the leaf-set $L(T)$ is $\Gen$ and a specific event-label is chosen
so that the least common ancestor of two distinct elements $x,y\in \Gen$ is labeled
in a way that uniquely determines whether $(x,y)\in R$ or not. That is, 
an event-labeled tree $(T,t)$ with events ``$0$'' and ``$1$'' on its inner vertices 
represents a (symmetric irreflexive) binary relation $R$ if for all $(x,y)\in \lfloor\Gen\times \Gen\rfloor_{\textrm{irr}}$
 it holds that $\lca_T(x,y) \ev 1$  if and only if $(x,y)\in R$. 

The latter definitions can easily be extended to arbitrary disjoint (irreflexive 
symmetric) relations
$R_1,\dots,R_k$ over $\Gen$: An edge-colored 
graph $G_{R_1,\dots,R_k}=(\Gen, E:=\cup_{i=1}^k E_{R_i})$
represents the relations $R_1,\dots,R_k$ if it holds that
 $(x,y)\in R_i$ if and only if
$\{x,y\}\in E$ and the edge $\{x,y\}$ is colored with ``$i$''.
Analogously, an event-labeled tree $(T,t)$  
with events ``$0$'' and ``$1,\dots, k$'' on its inner vertices represents
the relations $R_1,\dots,R_k$ if for all $(x,y)\in \lfloor\Gen\times \Gen\rfloor_{\textrm{irr}}$
 it holds that $\lca_T(x,y) \ev i$  if and only if $(x,y)\in R_i$, 
$1\leq i\leq k$. The latter implies that for all pairs $(x,y)$ that are in none of the relations
$R_i$ we have $\lca_T(x,y) \ev 0$. 

In practice, 
the disjoint relations correspond to the evolutionary relationship between
genes contained in $\Gen$, as e.g. the disjoint relations $R_{o}$ and $R_{p}$ that comprise
the pairs of orthologous and paralogous genes, respectively.

\paragraph{\textbf{Paralogy, Orthology, and Xenology}}

The current flood of genome sequencing data poses new challenges for
comparative genomics and phylogenetics. An important topic in this context
is the reconstruction of large families of homologous proteins, RNAs, and
other genetic elements. The distinction between orthologs, paralogs, and
xenologs is a key step in any research program of this type. The
distinction between orthologous and paralogous gene pairs dates back to the
1970s: 
two genes whose least common ancestor in the gene tree
corresponds to a duplication are paralogs; 
if the least common ancestor was
a speciation event and the genes are from different species, they are orthologs \cite{Fitch:70}. The importance
of this distinction is two-fold: On the one hand, it is informative in genome
annotation and, on the other hand, the orthology (or paralogy) relation
conveys information about the events corresponding to internal nodes of the
gene tree \cite{HHH+13} and about the underlying species tree
\cite{HLS+15,HHH+12}. 
We are aware of the controversy about the distinction between 
orthologous and paralogous genes and their consequence in the context of gene function, 
however, we adopt here the point of view that homology, and
therefore also orthology and paralogy, refer only to the evolutionary history of a
gene family and not to its function \cite{GB00, GK13}.

In contrast to orthology and paralogy, the definition of xenology is less well
established and by no means consistent in the biological literature. Xenology is
defined in terms of \emph{horizontal gene transfer (HGT)}, that refers to the
transfer of genes between organisms in a manner other than traditional
reproduction and across species. 
The most commonly used definition stipulates that two genes are
\emph{xenologs} if their history since their common ancestor involves
horizontal gene transfer of at least one of them \cite{Fitch2000,Jensen:01}.
In this setting, both
orthologs and paralogs may at the same time be xenologs \cite{Jensen:01}.
Importantly, the mathematical framework established for evolutionary ``event''-relations, 
as the orthology relation \cite{Boeckner:98,HHH+13}, naturally
accommodates more than two types of events associated with the internal nodes of
the gene tree. It is appealing, therefore, to think of a HGT event as
different from both speciation and duplication, in line with \cite{Gray:83}
where the term ``xenologous'' was originally introduced.

\begin{figure}[tbp]
  \begin{minipage}{.45\linewidth}
		\begin{itemize} 
			\item[] \scriptsize $R_{o} = \{dv\mid v\in \Gen\setminus\{d\}\}\cup \{ab_1, ac_2, b_1c_2, b_2c_2\}$
			\item[] $R_{x} = \{b_3c_1\}$
			\item[] $R_{p} = \lfloor\Gen\times \Gen\rfloor_{\textrm{irr}} \setminus (R_o\cup R_x)$ \\
			\item[] $xy\in R_{\star}$ means that  $(x,y)(y,x)\in R_{\star}$, with $\star \in \{o,p,x\}$ \\[0.1cm]
		\end{itemize}
		\centering
	  \includegraphics[scale = 0.35]{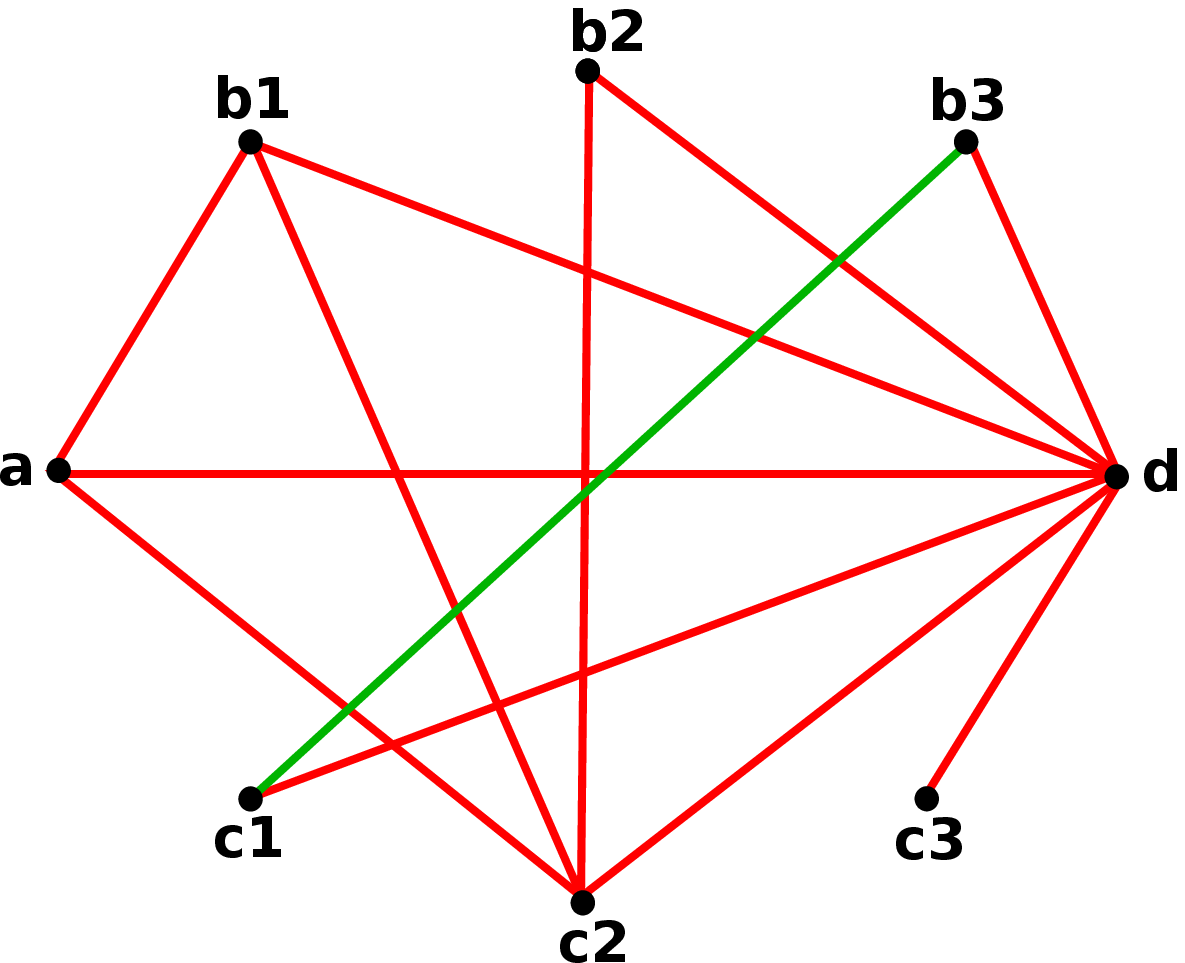} \\[0.5cm]  
  \end{minipage}
  \begin{minipage}{.4\linewidth}
  \includegraphics[scale = 0.45]{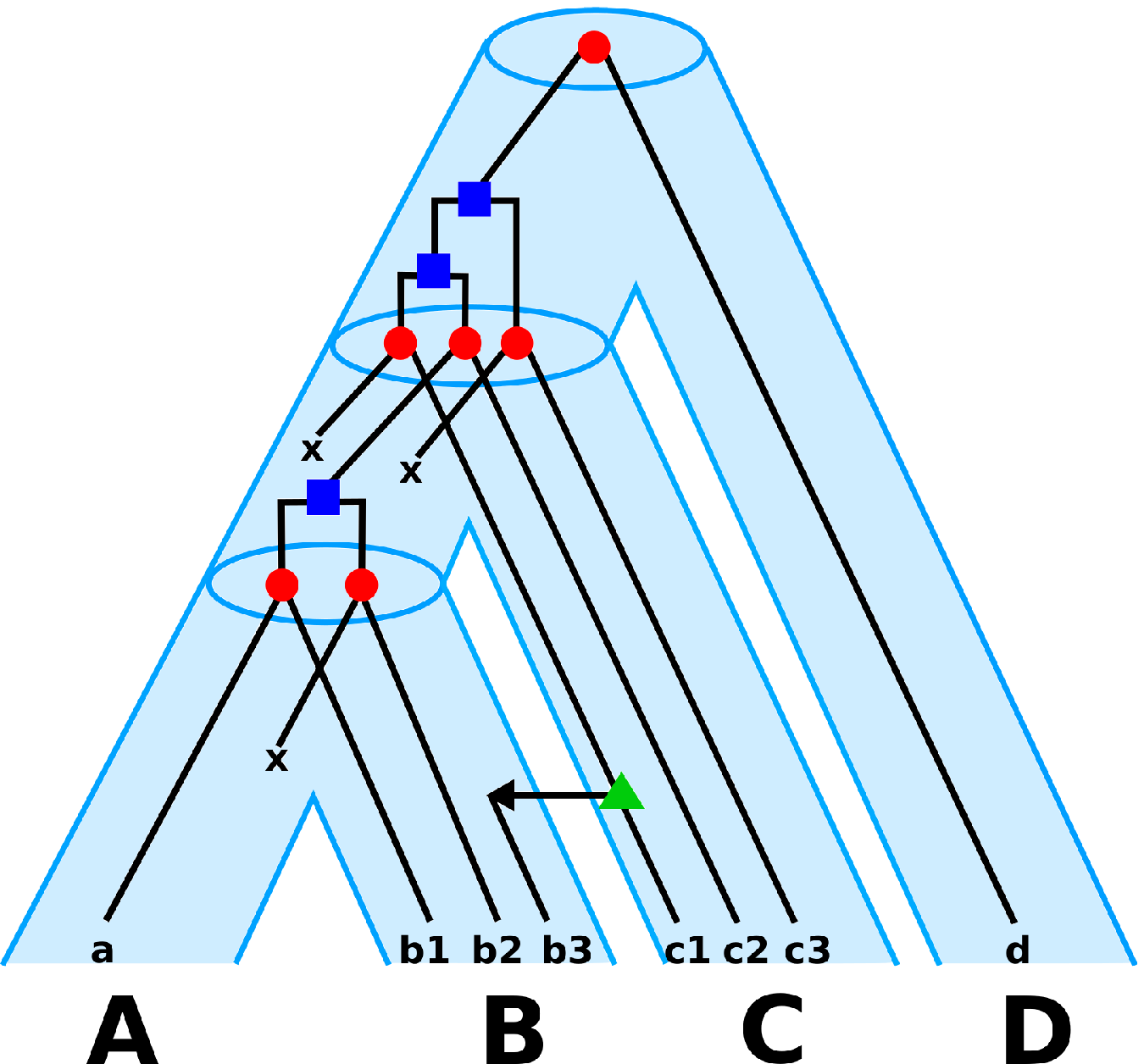} 
  \end{minipage}
  \caption{ Example of an evolutionary scenario showing the ``true'' 
				evolution of a gene family evolving along the species 
				tree (shown as blue tube-like tree).
            The corresponding true gene tree $T$ appears embedded in the species tree $S$. 
            The speciation vertices in the gene tree (red circuits) appear on the vertices of 
				the species tree (blue ovals), while the duplication
            vertices (blue squares) and the HGT-vertices (green triangles) 
		      are located on the edges of the species tree. 
				Gene losses are represented with ``{\textsf{x}}''. 
            The true gene-tree $T$ uniquely determines the relationships between the genes
				by means of the event at $\lca_T(x,y)$ of distinct genes 
				$x,y\in \Gen$.
				The pairs of $\lca$-orthologous, -paralogous and -xenologous genes  are comprised in
				the relations $R_o,R_p$ and $R_x$, respectively. 
				The graph-representation $G_{R_o,R_p,R_x}$ is shown in the lower left part. 
            Non-drawn edges indicate the paralogous genes. 
            This graph  clearly suggests that
				the orthology-relation $R_o$ is not a complete subgraph, and thus, \emph{does not}
            cluster or partition the input gene set $\Gen$. However, in all cases the subgraphs
            $G_{R_o}, G_{R_p}$ and $G_{R_x}$ are so-called cographs, cf.\ Thm.\ \ref{thm:charact}.
          }
  \label{fig:trueHist}
\end{figure}

In this contribution, we therefore will consider a slight modification of the terms
orthologs, paralogs and xenologs, so-called $\lca$-orthologs, $\lca$-paralogs
and $\lca$-xenologs. To this end, note that for a set of genes $\Gen$, 
the evolutionary relationship between two homologous genes contained in $\Gen$ 
is entirely explained by the \emph{true} evolutionary gene-history of these genes.
More precisely, if $T$ is a (known) tree reflecting the 
true gene-history together with the events that happened, that is, the labeling $t$ that 
tags the inner vertices of $T$ as a speciation, duplication or HGT event, 
respectively, then we can determine  the 
three disjoint relations $R_o, R_p$ and $R_x$ comprising
the pairs of so-called $\lca$-orthologous, $\lca$-paralogous and $\lca$-xenologous genes, respectively, 
as follows: Two genes $x,y\in \Gen$ are
\begin{itemize}
	\item \emph{$\lca$-orthologous}, if $\lca_T(x,y) \ev \text{speciation}$;
	\item \emph{$\lca$-paralogous},\  if $\lca_T(x,y) \ev \text{duplication}$ and 
	\item \emph{$\lca$-xenologous},\   if $\lca_T(x,y) \ev \textrm{HGT}$. 
\end{itemize}
The latter also implies the 
edge-colored graph representation $G_{R_o,R_p,R_x}$, see Figure \ref{fig:trueHist} for an
illustrative example.

In the absence of horizontal gene transfer, the relations 
$\lca$-orthologs and $\lca$-paralogs are equivalent to orthologs and
paralogs as defined by Fitch \cite{Fitch2000}. 

We are aware of the fact that this definition of $\lca$-``events'' leads to 
a loss of information of the direction of the HGT
event, i.e., the information of  donor and acceptor.
However, for the proposed method and to understand the idea
of representing estimates of evolutionary relationships
in an event-labeled tree this information is not 
necessarily needed. 
Nevertheless, generalizations to tree-representations of non-symmetric relations
or a mathematical framework for xenologs w.r.t.\ the notion of Fitch 
might improve the proposed methods. 

\begin{remark}
If there is no risk of confusion and  if not stated differently, 
we call $\lca$-orthologs, $\lca$-paralogs, and $\lca$-xenologs simply
orthologs, paralogs and xenologs, respectively.
\end{remark}

Clearly, evolutionary history and 
the events of the past cannot be observed directly and hence, must be inferred, using algorithmic and
statistical methods, from the genomic data available today.
Therefore, we can only deal with estimates of the relations $R_o, R_p$ and $R_x$. 
In this contribution, we use those estimates to reconstruct 
(a hypothesis of) the evolutionary history
of the genes and, eventually, the history of the species the genes reside in. 

We wish to emphasize that the three relations $R_o, R_p$ and
$R_x$ (will) serve as illustrative examples and the cases $R_p=\emptyset$
or $R_x=\emptyset$ are allowed. In practice, it is possible to have more
than these three relations. By way of example, the relation containing the pairs
of paralogous genes might be more refined, since gene duplications have
several different mechanistic causes that are also empirically distinguishable
in real data sets. Thus, instead of heaving a single relation $R_p$ that
comprises all paralogs, we could have different types of paralogy relations that 
distinguish between events
such as local segmental duplications, duplications by retrotransposition, or
whole-genome duplications \cite{Zhang:03}.

\section{From Sequence Data to Species Trees}
\label{sec:genTOspe}

In this section, we provide the main steps in order to infer event-labeled gene
trees and species trees from respective estimated event-relations. An
implementation of these steps by means of integer linear programming is provided
in the software tool \texttt{ParaPhylo} \cite{HLS+15}.

The starting point of this method is an estimate
of the (true) orthology relation $R_o$. From this estimate
the necessary information of the event-labeled gene trees and
the respective species trees will be derived. 

\subsection{Orthology Detection}

The inference of the orthology relation $R_o$ and lies at the heart of many reconstruction methods. 
Orthology inference methods can be classified based on the methodology they use
to infer orthology into \emph{tree-based} and \emph{graph-based} methods, for an overview 
see e.g.\ \cite{Altenhoff:09, DAAGD2013,gabaldon:08,KWMK:11,T+11}.

\emph{\textbf{Tree-based orthology inference methods}} rely on the
reconciliation of a constructed gene tree (without event-labeling) from an
alignment of homologous sequences and a given species tree, see e.g.\
\cite{Arvestad03072003,SPJ:11,Hubbard+07,HBNH:07,WPFR:07}. 

Although
tree-based approaches are often considered as very accurate given a species
tree, it suffers from high computational costs and is hence limited in practice
to a moderate number of species and genes. A further limitation of those
tree-reconciliation methods is that for many scenarios the species tree is not
known with confidence and, in addition, all practical issues that complicate
phylogenetic inference (e.g. variability of duplication rates, mistaken
homology, or HGT) limit the accuracy of both the gene and the species trees. 

Intriguingly, with \emph{\textbf{graph-based orthology inference methods}} it is 
possible in practice to detect the pairs of orthologous genes with acceptable accuracy \emph{without}
constructing either gene or species trees. Many tools
of this type have become available over the last decade. To name only a
few, 
\texttt{COG} \cite{TG+00}, \texttt{OMA} \cite{SDG:07,ASGD:11}, \texttt{eggNOG} \cite{JJK+08},
\texttt{OrthoMCL} \cite{li2003orthomcl, CMSR:06},  \texttt{InParanoid} \cite{inparanoid:10},
\texttt{Roundup 2.0} \cite{DeLuca12}, \texttt{EGM2} \cite{Mahmood30122011} or 
\texttt{ProteinOrtho} \cite{Lechner:11a} and its extension \texttt{PoFF}
\cite{Lechner:14}.
Graph-based methods detect 
orthologous genes for two (pairwise) or more (multiple) species. 
These methods consist of a \emph{graph construction phase} and, 
in some cases, a \emph{clustering phase} \cite{T+11}.
In the graph construction phase, a graph is inferred where vertices
represent genes, and (weighted) edges the (confidence of) orthology
relationships. 
The latter rely on pairwise sequence similarities (e.g., basic
local alignment search tool (BLAST) or Smith-Waterman) calculated between all
sequences involved and an operational definition of orthology,
for example, reciprocal best hit (RBH), bi-directional best hit (BBH), symmetrical best
hit (SymBeT) or reciprocal smallest distance (RSD). 
In the clustering phase, clusters or groups of orthologs are constructed, 
using e.g., single-linkage, complete-linkage, spectral clustering or 
Markov Cluster algorithm. 
However, orthology is a symmetric, but not a transitive relation, i.e.,
it does in general not represent a partition of the set of genes $\Gen$.
In particular, a set $\Gen'$ of genes can be orthologous to another gene
$g\in \Gen\setminus \Gen'$ but the genes within $\Gen'$ are not necessarily orthologous to each other. In this case, the genes in $\Gen'$ are
called co-orthologs to gene $g$ \cite{Koonin:05}. 
It is important to mention that, therefore, the problem of orthology detection is fundamentally different from clustering
or partitioning of the input gene set. 

In addition to \texttt{OMA} and \texttt{ProteinOrtho}
only \texttt{Synergy}, \texttt{EGM2}, and \texttt{InParanoid} 
attempt to resolve the orthology relation at the level of gene pairs.
The latter two tools can only be used for the analysis of two species at a time, while \texttt{Synergy} 
is not available as standalone tool and therefore cannot be applied to arbitrary user-defined data sets. 
In particular, the use of orthology inference tools is often limited to the
species offered through the databases published by their authors. An exception
is provided by \texttt{ProteinOrtho} \cite{Lechner:11a} and its extension
\texttt{PoFF} \cite{Lechner:14}, methods that we will use in our approach.
These standalone tools are specifically
designed to handle large-scale user-defined data and can be applied to hundreds
of species containing millions of proteins at ones. In particular, 
such computations can be performed on off-the-shelf hardware
\cite{Lechner:11a}. \texttt{ProteinOrtho} and \texttt{PoFF} compare similarities
of given gene sequences (the bit score of the blast alignment) that together
with an an E-value cutoff yield an edge-weighted directed graph. Based on
reciprocal best hits, an undirected subgraph is extracted (graph construction
phase) on which spectral clustering methods are applied (clustering phase), to
determine significant groups of orthologous genes. To enhance the prediction
accuracy, the relative order of genes (synteny) can be used as additional
feature for the discrimination between orthologs and paralogs. 

\smallskip
To summarize, graph-based methods have in common, that
the output is a set of (pairs of) putative orthologous genes. 
In addition, orthology detection tools often report some
weight or confidence value $w(x,y)$ for $x$ and $y$ 
to be orthologs or not. 
This gives rise to a symmetric, irreflexive binary relation 
\begin{align}
	\eT &= \{(x,y) \mid x,y \in \Gen \text{\ are\ estimated\ orthologs} \} \\
					 &=  \{(x,y) \mid \lca_T(x,y) \ev \textrm{speciation} \text{\ (in the estimated gene tree }T\text{)} \}.
\end{align}

\subsection{Construction of Gene Trees}

\paragraph{\textbf{Characterization of Evolutionary Event Relations}}

Assume we have given a ``true'' orthology relation $R_o$ over $\Gen$,  i.e., 
$R_o$ comprises all pairs of ``true'' orthologs, that is, if the true evolutionary history  $(T,t)$ of the
genes would be known, then $(x,y) \in R_o$ if and only if $\lca_T(x,y) \ev \text{\
speciation}$. 
As we will show, given such a  true relation without the knowledge of the gene
tree $(T,t)$, it is possible to reconstruct the ``observable discriminating part'' of $(T,t)$ using
the information contained in $R_o$, resp., $R_p$ only, at least in the absence of xenologous
genes \cite{HHH+13,HW:15}.  In the presence of HGT-events, but given the ``true'' relations
 $R_o$ and $R_p$ it is even possible to reconstruct $(T,t)$ using
the information contained in $R_o$ and $R_p$ only \cite{HHH+13,HW:15}. Note, for the set of pairs of
($\lca$-)xenologs $R_x$ we have
\[R_x = \lfloor\Gen\times \Gen\rfloor_{\textrm{irr}}\setminus (R_o\cup R_p ).\]

Clearly, since we do not know the true evolutionary history with
confidence, we always deal with estimates $\eT, \eP, \eX$ of these true relations $R_o, R_p, R_x$. 
In order to understand under which conditions it is possible 
to infer a gene tree $(T,t)$ that represents the disjoint estimates
$\eT, \eP, \eX$, 
we characterize in the following the structure of their graph-representation $G_{\eT,\eP,\eX}$. 
Note, if 
$\eT\cup \eP\cup \eX = \lfloor\Gen\times \Gen\rfloor_{\textrm{irr}}$, then $G_{\eT,\eP,\eX}$ is a complete edge-colored graph, i.e., 
for all distinct $x,y\in \Gen$ there is an edge $\{x,y\}\in E$ s.t.\ 
$\{x,y\}$ is colored with with ``$\star$'' if and only if $(x,y)\in R_{\star}$, 
$\star \in \{o,p,x\}$

The following theorem is based on results established by B\"{o}cker and Dress \cite{Boeckner:98} and 
Hellmuth et al.\ \cite{HHH+13}. 
\begin{theorem}[\cite{Boeckner:98,HHH+13}]
    Let $G_{R_1,\dots,R_k}$ be the graph-representation of the relations 
    $R_1,\dots,R_k$ over some set $\Gen$.
	 There is an event-labeled gene tree representing $R_1,\dots,R_k$ if and only if
	 \begin{itemize}
			\item[(i)]\ the graph $G_{R_i}=(\Gen, E_{R_i})$  is a 
                    cograph for all $i\in\{1,\dots,k\}$ and
         \item[(ii)]\ for all three distinct genes $x,y,z\in \Gen$ the three
                  	edges $\{x,y\}, \{x,z\}$ and $\{y,z\}$ in $G_{R_1,\dots,R_k}$ 
                     have at most two distinct colors. 
	 \end{itemize}
	\label{thm:charact}
\end{theorem}

Clearly, in the absence of xenologs and thus, if $\eP = \lfloor\Gen\times \Gen\rfloor_{\textrm{irr}} \setminus \eT$, 
we can ignore condition $(ii)$, since at most two colors occur in $G_{\eT,\eP}$. 
In the latter case, $G_{\eT}$, resp., $G_{\eP}$
alone provide all information of the underlying gene tree.

Theorem \ref{thm:charact} implies that whenever we have estimates $\eT$ or $\eP, \eX$
and we want to find a tree $(T,t)$ that represents these relations we must
ensure that neither $G_{\eT}$, $G_{\eP}$ nor $G_{\eX}$ contains an induced path
on four vertices and that there is no triangle (a cycle on three vertices) in
$G_{\eT,\eP,\eX}$ where each edge is colored differently. However, due to noise
in the data or mispredicted events of pairs of genes, the graph
$G_{\eT,\eP,\eX}$ will usually violate condition $(i)$ or $(ii)$. A particular
difficulty arises from the fact, that we usually deal with the estimate $\eT$ only,
and do not know how to distinguish between the paralogs and xenologs.

One possibility to correct the initial estimates $\eT,\eP, \eX$ to the ``closest''
relations $R^*_o,R^*_p, R^*_x$ so that there is a tree representation of
$R^*_o,R^*_p, R^*_x$, therefore, could be the change of a minimum number of edge-colors in 
$G_{\eT,\eP,\eX}$ so that $G_{R^*_o,R^*_p,
R^*_x}$ fulfills Condition $(i)$ and $(ii)$. This problem was recently
shown to be NP-complete \cite{HW:15,HW16,Liu:12}.

\paragraph{\textbf{Inference of Local Substructures of the Gene Tree}}

Assume we have given (estimated or true) relations $R_o,R_p, R_x$ so that 
the graph-representation $G_{R_o, R_p, R_x}$ fulfills Condition $(i)$
and $(ii)$ of Theorem \ref{thm:charact}. We show now briefly, how to construct
the tree $(T,t)$ that represents $R_o, R_p, R_x$. 

Here we utilize the information of triples that are extracted from the graph 
$G_{R_o, R_p, R_x}$ and that must be contained in any gene-tree $(T,t)$ representing 
$R_o, R_p, R_x$.
More precisely, given the relations $R_1,\dots,R_k$ we define the set of 
triples $\mathcal{T}_{R_1,\dots,R_k}$ as follows:
For all three distinct genes $x,y,z\in \Gen$ we add
the triple $\rt{xy|z}$ to $\mathcal{T}_{R_1,\dots,R_k}$ if and only if the colors of the edge 
$\{y,z\}$ and $\{x,z\}$ are identical but distinct from the color of the edge 
$\{x,y\}$ in $G_{R_1,\dots,R_k}$.
In other words, for the given evolutionary relations $R_o, R_p, R_x$ the triple 
$\rt{xy|z}$ is added to $\mathcal{T}_{R_o, R_p, R_x}$ iff the two genes $x$ and $z$, as well as $y$ and $z$ 
are in the same evolutionary relationship, but different from the evolutionary
relation between $x$ and $y$.

\begin{theorem}[\cite{Boeckner:98,HHH+13}]
    Let $G_{R_1,\dots,R_k}$ be the graph-representation of the relations 
    $R_1,\dots,R_k$.
    The graph $G_{R_1,\dots,R_k}$ fulfills conditions $(i)$ and $(ii)$	
	 of Theorem \ref{thm:charact} (and thus, there is a tree representation of 
    $R_1,\dots,R_k$) 
  	 if and only if there is a tree $T$ that contains all the triples 
    in $\mathcal{T}_{R_1,\dots,R_k}$.
\end{theorem}

The importance of the latter theorem lies in the fact, that 
the well-known algorithm \texttt{BUILD} \cite{Aho:81,sem-ste-03a} can be applied to
$\mathcal{T}_{R_1,\dots,R_k}$ to 
determine whether the set of triples $\mathcal{T}_{R_1,\dots,R_k}$  is consistent, 
and, if so, 
constructs a tree representation in polynomial-time. To obtain a valid
event-label for such a tree $T$ we can simply set
$t(\lca_T(x,y)) = \star$ if the color of the edge $\{x,y\}$ in 
$G_{R_o, R_p, R_x}$ is ``$\star$'', $\star\in \{o,p,x\}$ \cite{HHH+13}.

\begin{figure}[tbp]
	\centering
  \includegraphics[scale = 0.65]{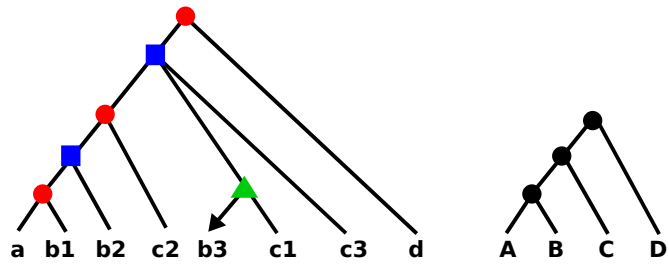} \\[0.1cm] 
  \caption{Left, the homeomorphic image $(T,t)$ of the observable gene tree  $(T',t')$ in Fig.\ \ref{fig:trueHist} is shown. 
           The true gene tree in Fig.\ \ref{fig:trueHist} represents all extant as well 
           extinct genes, all duplication, HGT and speciation events.
           Not all of these events are observable from extant genes
           data, however. 
			  In particular, extinct genes cannot be observed. 
			  Thus, the observable gene tree  $(T',t')$ is obtained from the original gene tree in Fig.\ \ref{fig:trueHist}
			   by removing
           all vertices marked with ``{\textsf{x}}'' together with their incident edges and, thereafter, 
			  removing all inner vertices that are contained in only two edges.
           The homeomorphic image $(T,t)$ is obtained from $(T',t')$ 
           by contraction of the edge that connect the two consecutive duplication events. 
          	The species triple set $\mathcal S$ is $\{\rt{AB|C}_2, \rt{AB|D}_3, \rt{AC|D}_3, \rt{BC|D}_9 \}$, 
				where indices indicate the number of gene triples in $\mathcal R_o(T)$ that support the 
				respective species triple. 
				In this example, the (unique, and thus minimally resolved) species tree $S$
				that contains all triples in $\mathcal S$ is shown in the right part.
            The species tree $S$ is identical to the true species tree shown in Fig.\ \ref{fig:trueHist}
          }
  \label{fig:obs-trueHist}
\end{figure}

It should be stressed that the evolutionary relations do not contain 
the full information on the event-labeled gene tree, see Fig. \ref{fig:obs-trueHist}. 
Instead, the constructed  gene trees $(T,t)$ are homeomorphic images of the 
(possibly true) observable gene tree $(T',t')$ by collapsing adjacent events 
of the same type \cite{HHH+13}. That is, in the  constructed tree $(T,t)$ all
inner vertices that are connected by an edge will have different event-labels, 
see Fig \ref{fig:obs-trueHist}.
Those trees are also known as discriminating representation, cf.\ \cite{Boeckner:98}.
However, these discriminating representations contain and provide  the necessary information
to recover the input-relations, are unique (up to isomorphism), and do not
pretend a higher resolution than actually supported by the data.

\subsection{Construction of Species Trees}

While the latter results have been established for \mbox{($\lca$)-orthologs}, \mbox{-paralogs} and 
\mbox{-xenologs}, we restrict our attention in this subsection to orthologous
and paralogous genes only and assume that there are no
HGT-events in the gene trees. We shall see later, that  in practical
computation the existence of xenologous genes does not have a large impact 
on the reconstructed species history, although the theoretical results
are established for gene histories without xenologous genes.

In order to derive for a gene-tree $(T,t)$ (that contains only speciation and duplication
events) a species tree $S$ with which $(T,t)$
can be reconciled with or simply spoken ``embedded'' into, we need to answer the question
under which conditions there exists such a species tree for a given gene tree. 

A tree $S=(W,F)$ with leaf set $\Spe$ is a species tree for a gene tree
$T=(V,E)$ with leaf set $\Gen$ if there is a reconciliation map $\mu:V\to W\cup F$
that map the vertices in $V$ to vertices or edges in $W\cup F$. 
A reconciliation map $\mu$ maps the genes $x\in \Gen$ in $T$ to the respective 
species  $X \in \Spe$ in $S$ the gene $x$ resides in so that specific constraints 
are fulfilled. In particular, the inner vertices of $T$ with label ``speciation''
are mapped to the  inner vertices of $S$, while the duplication vertices of
$T$ are mapped to the edges in $W$ so that the relative ``evolutionary order'' of
the vertices in $T$ is preserved in $S$. We refer to \cite{HHH+12} for the full definition
of reconciliation maps. In Fig.\ \ref{fig:trueHist}, the 
reconciliation map $\mu$ is implicitly given by drawing the species tree superimposed on the gene tree.

Hence, for a given gene tree $(T,t)$ we wish to efficiently decide whether there is
a species tree in which $(T,t)$ can be embedded into, and if so, construct such a
species tree together with the respective reconciliation map. 
We will approach the problem of 
deriving a species tree from an event-labeled gene tree
by reducing the reconciliation map from gene tree to
species tree to rooted triples of genes residing in three distinct species. 
To this end we define a species triple set $\mathcal S$ derived from $(T,t)$ 
that provides all information
needed to efficiently decide whether there is a species tree $S$ for $(T,t)$  or not.

Let $\mathcal{R}_o(T)$ be the set of all triples $\rt{ab|c}$ that are contained in $T$ s.t.\	 
$a,b,c\in \Gen$ reside in pairwise different species
and $\lca_T(a,b,c)\ev \text{speciation}$, then set
\[\mathcal S \coloneqq \{\rt{AB|C} \colon \exists \rt{ab|c}\in \mathcal R_o(T) \text{ with } a\in A, b\in B, c\in C\}.\]
It should be noted that by results established in \cite{Boeckner:98, HHH+13} 
it is possible to derive the triple set $\mathcal S$ directly from the orthology relation
$R_o$ without constructing a gene tree, cf.\ \cite{HHH+13}:
$\rt{AB|C}\in \mathcal S$ if and only if 
\begin{enumerate}
 \item[(I)] $A,B$ and $C$ are pairwise different species
\end{enumerate}
\emph{and} there  are genes $a\in A, b\in B, c\in C$ so that \emph{either} 
\begin{enumerate}
 \item[(IIa)]$(a,c),(b,c)\in R_o$ and $(a,b)\not\in R_o$ \emph{or}
 \item[(IIb)]$(a,c),(b,c),(a,b)\in R_o$ and there is a gene $d\in \Gen$
               with $(c,d)\in R_o$ and $(a,d),(b,d)\notin R_o$. 
\end{enumerate}
Thus, in order to infer species triples a sufficient number of 
duplication events must have happened. 
The following important result was given in \cite{HHH+12}. 
\begin{theorem}
	Let $(T,t)$ be a given gene tree that contains only speciation and duplication events. 
	Then there is a species tree $S$ for $(T,t)$ if and only if there is any tree
   containing all triples in $\mathcal S$.

   In the positive case, the species tree $S$ and the reconciliation between $(T,t)$ and $S$ 
   can be found in polynomial time. 
   \label{thm:reconc}
\end{theorem}

Interestingly, the latter theorem implies that 
the gene tree $(T,t)$ can be embedded into any tree that contains
the triples in $\mathcal S$. Hence, one usually wants to find a species 
tree with a least number of inner vertices, as those trees
constitute one of the best estimates of the phylogeny
without pretending a higher resolution than actually supported by the
data. Such trees are also called minimally resolved tree and computing such
trees is an NP-hard problem \cite{Jansson:12}.

Despite the variance reduction due to cograph editing, noise in the data,
as well as the occasional introduction of contradictory triples as a
consequence of horizontal gene transfer is unavoidable. The species triple set
$\mathcal S$ 	collected from the individual gene families thus will not always be
consistent.
The problem of determining a
maximum consistent subset of an inconsistent set of triples is
NP-hard and also APX-hard, see \cite{Byrka:10a,vanIersel:09}. Polynomial-time
approximation algorithms for this problem and further theoretical results
are reviewed in	 \cite{Byrka:10}.

The results in this subsection have been established for the reconciliation 
between event-labeled gene trees \emph{without} HGT-events and 
inferred species. 
Although there are reconciliation maps defined for gene trees that contain xenologs 
and respective species trees \cite{Bansal-HGT, Bansal-HGT2}, 
a mathematical characterization of the species triples $\mathcal S$ and the existence 
of species trees for those gene trees, which might help also 
to understand the transfer events itself, however, is still an open problem.

\subsection{Summary of the Theory}
The latter results show that it is not necessary to restrict the inference
of species trees to 1:1 orthologs. 
Importantly, orthology information alone is
sufficient to reconstruct the species tree provided that (i) the
orthology is known without error and unperturbed by horizontal
gene transfer and (ii) the input data contains a sufficient number
of duplication events. Although species trees can be inferred
in polynomial time for noise-free data, in a realistic setting, three
NP-hard optimization problems need to be solved.

We summarize the important working steps to infer
the respective gene and species trees  from genetic material.

\begin{itemize}
	\item[(W1)]$\quad$ Compute the estimate $\eT$ and set 
    			  $\eP = \lfloor\Gen\times \Gen\rfloor_{\textrm{irr}}\setminus \eT$.
	\item[(W2)]$\quad$ Edit the graph $G_{\eT}$ to the closest cograph with a minimum number of edge edits to
              obtain the graph  $G_{R_o}$. Note, $R_p = \lfloor\Gen\times \Gen\rfloor_{\textrm{irr}}\setminus R_o$.  
	\item[(W3)]$\quad$ Compute the tree representation $(T,t)$ w.r.t.\ $R_o, R_p$.  
	\item[(W4)]$\quad$ Extract the species triple set $\widehat{\mathcal S}$ from $\mathcal R_o(T)$. 
	\item[(W5)]$\quad$ Extract a maximal consistent triple set $\mathcal S$ from $\widehat{\mathcal S}$.
	\item[(W6)]$\quad$ Compute a minimally resolved species tree $S$ that contains all triples in $\mathcal S$, 
				  and, if desired, the reconciliation map $\mu$  between $(T,t)$ and $S$ 
					(cf.\ Thm.\ \ref{thm:reconc}).
\end{itemize}

In the presence of horizontal transfer, 	
in Step (W1) the xenologous genes $x,y$ are either predicted as orthologs
or paralogs.

Furthermore, in Step (W2) it suffices to edit the graph
$G_{\eT}$ only, since afterwards the graph representation $G_{R_p}$ with $R_p =
\lfloor\Gen\times \Gen\rfloor_{\textrm{irr}}\setminus R_o$ and, thus 
$G_{R_o,R_p}$ fulfills the conditions of Thm.\ \ref{thm:charact} \cite{Corneil:81,HHH+13}. 
In particular, 
the graphs $G_{R_o,R_p}$ and $G_{R_p}$ have then been obtained from
$G_{\eT,\eP}$, resp., $G_{\eP}$  with a minimum number of edge
edits. The latter is due to the fact that the complement $\overline G_{\eT}$ is the
graph $G_{\eP}$ \cite{Corneil:81}. 

To extract the species triple set $\widehat{\mathcal S}$ in Step (W4), 
it suffices to choose the respective species
triples using Condition (I) and (IIa)/(IIb), without constructing the gene trees and thus, 
Step (W3) can be ignored if the gene history is not of further interest.

\section{Evaluation}

In \cite{HLS+15} it was already shown that for real-life data sets
the paralogy-based method produces phylogenetic trees for moderately sized species
sets. The resulting species trees are comparable to those presented in the literature
that are constructed by ``state-of-the-art'' phylogenetic
reconciliation approaches as \texttt{RAxML} \cite{raxml:14}	 
or \texttt{MrBayes} \cite{mrbayes:12}.
To this end, genomic sequences of eleven \emph{Aquificales}
and 19 \emph{Enterobacteriales} species
were analyzed.
Based on the NCBI gene annotations of those species,
an orthology prediction was performed using \texttt{ProteinOrtho}.
From that prediction, phylogenetic trees were constructed
using the aforementioned orthology-paralogy-based approach (working steps (W2)-(W6)) implemented
as integer linear program in \texttt{ParaPhylo} \cite{HLS+15}.
The advantage of this approach is the computation of exact solutions, 
however, the runtime scales exponentially with the number of input genes per 
gene family and the number of species.

\begin{figure}[htbp]
    \centering
    \begin{subfigure}[b]{1\textwidth}
		  \centering
        \includegraphics[height = 3.3cm]{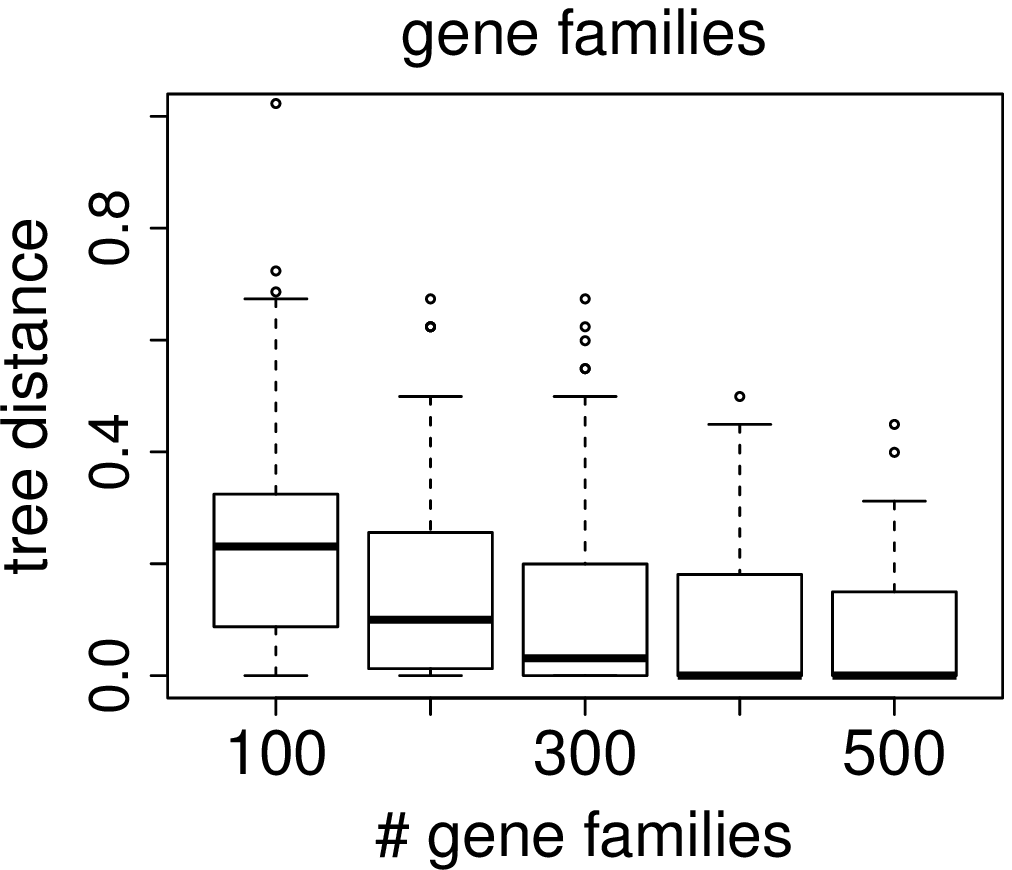}  \\ \
    \end{subfigure}
    \begin{subfigure}[b]{1\textwidth}		  \centering
        \includegraphics[height = 3.3cm]{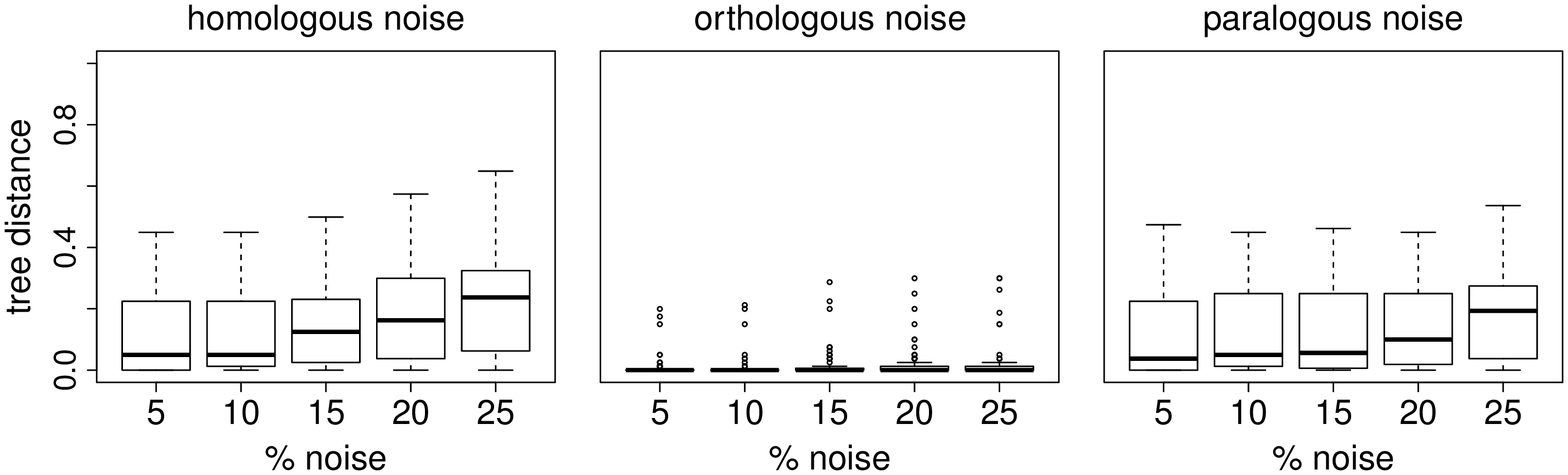}  \\	
    \end{subfigure}
    \begin{subfigure}[b]{1\textwidth}		  \centering
     \includegraphics[height = 3.65cm]{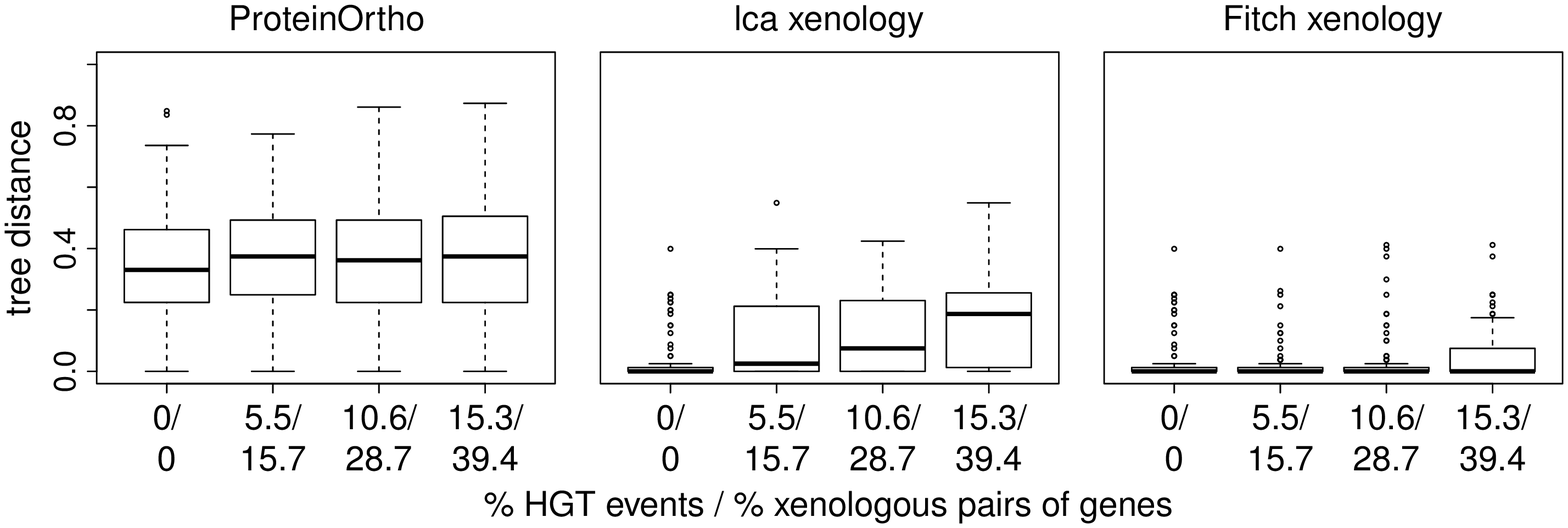} 
    \end{subfigure}
    \caption{Accuracy of reconstructed species trees (10 species) in simulated data sets; \emph{(Top)} 
				Dependence on the number of gene families; 
				\emph{(Middle)} Dependence of different noise models; \emph{(Down)} Dependence on noise by HGT.}
	\label{fig:boxplots}
\end{figure}

However, as there is no gold standard for phylogenetic tree reconstruction,
three simulation studies are carried out to evaluate the robustness of the method.
Using the \emph{Artificial Life Framework (ALF)} \cite{DAGD:12},
the evolution of generated gene sequences was simulated along a given branch
length-annotated species tree, explicitly taking into account
gene duplication, gene loss, and horizontal transfer events.
For realistic species trees, the $\gamma$-proteobacteria tree
from the OMA project \cite{ASGD:11} was randomly pruned to a size
of 10 species while conserving the branch lengths.
For additional details on the simulation see \cite{HLS+15}.
The reconstructed trees are then compared with the initial species
trees, using the software \texttt{TreeCmp} \cite{BGW-treeCmp:12}.
In the provided box-blots (Fig.\ \ref{fig:boxplots}), tree distances are computed according to the triple metric and normalized
by the average distance between random Yule trees, see \cite{HLS+15} for further evaluations.

The three simulation studies are intended to answer three individual questions.
\begin{enumerate}
  \item How much data is needed to provide enough information to reconstruct accurate species trees? (cf.\ Fig.\ \ref{fig:boxplots} (top))
  \item How does the method perform with noisy data?  (cf.\ Fig.\ \ref{fig:boxplots} (middle))
  \item What is the impact of horizontal gene transfer on the accuracy of the method?  (cf.\ Fig.\ \ref{fig:boxplots} (down))
\end{enumerate}

To construct accurate species trees, the presented method requires a sufficient amount of
duplicated genes.
Assuming a certain gene duplication rate, the amount of duplicated genes
correlates directly with the number of genes per species, respectively the number of gene families.
The first simulation study (Fig.\ \ref{fig:boxplots} (top)) is therefore performed with several numbers of gene families,
varying from 100 to 500. The simulation with \emph{ALF} was performed without horizontal gene transfer
and the phylogenetic trees are computed based on the unaltered orthology/paralogy relation
obtained from the simulation, that is, the orthologs and paralogs  
can directly be derived from the simulated gene trees. 
It turned out that with an duplication rate of 0.005, which corresponds to approximately 8\%
of paralogous pairs of genes, 500 gene families are sufficient to produce reliable phylogenetic trees.
With less gene families, and hence less duplicated genes, the trees tend to be only poorly resolved.

For the second study the simulated orthology/paralogy relation of 1000 gene families
was perturbed by different types of noise.
(i) insertion and deletion of edges in the orthology graph (homologous noise),
(ii) insertion of edges (orthologous noise),
and
(iii) deletion of edges (paralogous noise), see Fig.\ \ref{fig:boxplots} (middle).
In the 
three models an edge is inserted or removed with probability $p \in \{0.05, 0.1, 0.15, 0.2, 0.25\}$.
It can be observed that up to noise of approximately 10\%
the method produces trees which are almost identical to the initial trees.
Especially, in the case of orthology overprediction (orthologous noise)
the method is robust even if 25\% of the input data was disturbed.

Finally, in the third analysis, data sets are simulated with different rates of
horizontal gene transfer, see Fig.\ \ref{fig:boxplots} (down). The number of HGT
events in the gene trees are varied up to 15.3\%, which corresponds to 39.4\% of
all pairs of genes $(x,y)$ having at least one HGT event on the path from $x$ to
$y$ in the generated gene tree, i.e., $x$ and $y$ are xenologous with respect to
the definition of Fitch \cite{Fitch2000}. Firstly, the simulated gene
sequences are analyzed using \texttt{ProteinOrtho} and the tree reconstruction
is then performed based on the resulting orthology/paralogy prediction (Fig.\
\ref{fig:boxplots} (down/left)). Secondly, we used both definitions of xenology,
i.e., $\lca$-xenologous and the notion of Fitch. Note, so far the reconstruction
of species trees with \texttt{ParaPhylo} requires that pairs of genes are either
orthologous or paralogous. Hence, we used the information of the
$\lca$-orthologs, -paralogs and -xenologs derived from the simulated gene
trees. Fig.\ \ref{fig:boxplots} (down/center) shows the accuracy of
reconstructed species trees under the assumption that all $\lca$-xenologs are
``mispredicted'' as $\lca$-orthologs, in which case all paralogous genes are
identified correctly. Fig.\ \ref{fig:boxplots} (down/right) shows the accuracy
of reconstructed species trees under the assumption that all xenologs w.r.t.\
the notion of Fitch are interpreted as $\lca$-orthologs. The latter amounts to
the ``misprediction'' of $\lca$-xenologs and $\lca$-paralogs, as 
$\lca$-orthologs. However, all remaining $\lca$-paralogs, are still correctly
identified. For the orthology/paralogy prediction based on
\texttt{ProteinOrtho}, it turned out that the resulting trees have a distance of
approximately 0.3 to 0.4 to the initial species tree. Thereby, a distance of 1
refers not to a maximal distance, but to the average distance between random
trees. However, the accuracy of the constructed trees appears to be independent
from the amount of horizontal gene transfer. Hence, \texttt{ProteinOrtho} is not
able to either identify the gene families correctly, or 
mispredicts orthologs and paralogs (due to, e.g., gene loss). 
In case that all paralogous genes are
identified correctly, \texttt{ParaPhylo} produces more accurate trees. We
obtain even more accurate species trees, when predicting all pairs of
Fitch-xenologous genes as $\lca$-orthologs, even with a large amount of HGT events.

\section{Concluding Remarks}
The restriction to 1:1 orthologs for the reconstruction
of the evolutionary history of species is not necessary. Even more, it has been
shown that the knowledge of only a few correct identified paralogs allows to
reconstruct accurate species trees, even in the presence of horizontal gene
transfer. The information of paralogs is strictly complementary to the sources
of information used in phylogenomics studies, which are always based on
alignments of orthologous sequences. Hence, paralogs contain meaningful and
valuable information about the gene and the species trees. Future research might
therefore focus on improvements of orthology and paralogy infence tools, and
mathematical frameworks for tree-representations of non-symmetric relations
(since HGT is naturally a \emph{directed} event), as well as a characterization
of the reconciliation between gene and species trees in the presence of HGT.

\bibliographystyle{plain}
\bibliography{biblio}
\end{document}